

\def\ds{\displaystyle}
\def\bxp{{{\bf x}^{\prime}}}
\def\bx{{\bf x}}
\def\by{{\bf y}}
\def\br{{\bf r}}
\def\bp{{\bf p}}

\def\oneha{{1\over 2}}
\def\aln{{\left ( {\alpha \over n} \right )}}
\def\NT{{Nieuwenhuis and Tjon }}
\def\kem{{{1\over m} \nabla^2}}
\def\kemm{{{1\over {2m}} \nabla^2}}
\def\phid{\phi^{\dag}}
\def\xp{x^\prime}

\def\di{\partial}
\def\ni{\noindent}

\def\phib{\overline \phi}

\baselineskip=20pt
\centerline{Some exact solutions of reduced scalar Yukawa theory}
\par\vskip 1.5truecm
\centerline{Jurij W. Darewych$^{\dag}$}
\par
\centerline{Department of Physics and Astronomy}
\par
\centerline{York University}
\par
\centerline{Toronto, Ontario}
\par
\centerline{M3J 1P3 Canada}
\par
\vskip 2truecm
$^{\dag}$ e-mail: darewych@yorku.ca
\vskip 1truecm
\ni PACS: 11.10.Ef, 11.10.Qr, 03.70.+k
\vskip 2truecm   
{\ni \bf Abstract}
\par\medskip
The scalar Yukawa model, in which a complex scalar field, $\varphi$, interact
  via a real scalar field, $\chi$,  is reduced by using covariant Green functions. 
It is shown that exact few-particle eigenstates of the truncated QFT Hamiltonian 
can be obtained  in the Feshbach-Villars formulation if an unorthodox ``empty''
 vacuum state is used.  
Analytic solutions for the two-body case are obtained for massless chion 
exchange in 3+1 dimensions and for massive chion exchange 
in 1+1 dimensions. Comparison is made to ladder Bethe-Salpeter, Feynman-Schwinger 
and quasipotential results for massive chion exchange in 3+1. Equations for the
three-body case are also obtained.

\vfill\eject

{\ni \bf 1. Introduction}
\vskip .5truecm
The model scalar field theory, based on the Lagrangian  
density ($\hbar = c = 1$)
$$\eqalignno {
{\cal L}&=\di^\nu \varphi^*(x)\, \di_\nu \varphi (x) - 
m^2 \varphi^*(x)  \varphi (x) \cr
&+ \oneha \di^\nu \chi (x) \,\di_\nu \chi (x) - \oneha 
\mu^2 \chi^2 (x)  - g \varphi^*(x) \varphi (x)\, \chi (x) & (1)\cr
}
$$
is often used as a prototype QFT in many  studies. In 
particular, for the case $\mu = 0$, it is known as the 
Wick-Cutkosky model [1,2]. This model has been investigated 
in various formalisms, in addition to the original 
ladder Bethe-Salpeter formulation [3], including the 
light-cone formulation [4,5], and others (see refs. 
[6-9] and [18, 20-23] and citations therein).

  We shall consider a reduced form of this theory in the present 
  paper, in which the mediating chion field is partially 
  eliminated by means of covariant Green functions, in a 
  manner analogous to that discussed recently for QED [10,11]. 
  In addition, we shall use the  Feshbach-Villars (FV)  
  formulation [12] for the complex phion field, and an unconventional 
 ``empty'' vacuum state,   as has been used recently to determine  solutions 
  of the $\lambda (\varphi^* \varphi)^2$ theory [13]. 

The fields $\varphi$ and $\chi$ of the model (1) satisfy the equations    
$$
\di^\nu \di_\nu \chi (x)+ \mu^2 \chi (x) = -g  \varphi^*(x) \varphi (x),
 \eqno (2)
$$
$$
\di^\nu \di_\nu \varphi (x) + m^2 \varphi (x) = -g  \varphi (x) \chi (x),
 \eqno (3)
$$
and its conjugate.

Equation (2) has the formal solution
$$
\chi(x) = \chi_0(x) + \int d\xp \,D(x-\xp)\, \rho(\xp), \eqno (4)
$$
where $dx = d^Nx \,dt$ in $N+1$ dimensions,  $\rho(x) = - g 
\varphi^*(x) \varphi (x)$, $\chi_0 (x)$ satisfies the homogeneous 
(or free field) equation (eq. (2) with $g = 0$), while $D(x-\xp)$ 
is a covariant Green function (or chion  propagator, in QFTheoretic 
language), such that
$$
\left ( \di^\nu \di_\nu  + \mu^2 \right ) D(x-\xp) = \delta^{N+1}(x-\xp).
 \eqno (5)
$$
Equation (5) does not specify $D(x-\xp)$ uniquely since, for 
example, any solution of the homogeneous equation can be added 
to it without invalidating (5). Boundary conditions based on 
physical considerations are used to pin down the form of $D$.

Substitution of the formal solution (4) into eq. (3) yields the 
equation
$$
\di^\nu \di_\nu \varphi(x) + m^2 \varphi(x) = -g  \varphi (x) 
\chi_0 (x) - g \varphi(x) \int d\xp D(x-\xp) \rho(\xp).
 \eqno (6)
$$
Equation (6) is derivable from the action principle $\ds \delta 
\int dx\, {\cal L} = 0$, corresponding to the Lagrangian density
$$\eqalignno {
{\cal L}=&\di^\nu \varphi^*(x) \di_\nu \varphi (x) - m^2 \varphi^*(x)  
\varphi (x) 
  - g \varphi^*(x) \varphi (x)\, \chi_0 (x) & (7)\cr
&+ \oneha \int d\xp \rho (x) D(x-\xp) \rho(\xp),
}
$$
provided that $D(x-\xp) = D(\xp-x)$. 

The QFTs based on (1) and (7) are equivalent in that, in conventional 
covariant perturbation theory, they lead to the same invariant 
matrix elements in various order of perturbation theory.  
The difference is that, in the formulation based on (7), the 
interaction term that contains the propagator corresponds to Feynman 
diagrams  involving virtual chions, while the  
term that contains $\chi_0$  correspond to diagrams that cannot be generated 
using the term with $D(x-\xp)$, such as those with external (physical) chion lines.

We shall consider this scalar theory in the Feshbach-Villars (FV) 
formulation [12].  The reason for doing so is that this leads to 
a QFTheoretic Hamiltonian which is Schr\"odinger-like in form, 
for which exact eigensolutions can be readily written down. 
In the FV formulation,  the  field $\varphi$ and its time-derivative 
$\dot \varphi$ are replaced by a two-component vector
$$
\phi=\left[\matrix{u = {1\over {\sqrt{2m}}}(m \varphi + i \dot \varphi)&\cr
v = {1\over {\sqrt{2m}}}( m \varphi - i \dot \varphi)&\cr}\right], \eqno (8)
$$
so that, for example, $2 m \,\varphi^* \varphi = (u^* + v^*)(u+v) 
= \phi^{\dag} \eta \tau \phi$, where $\eta$ and $\tau$ are the matrices
$$\eta = \left [\matrix{1&0\cr 0& -1\cr}\right ]\;\;\;\; {\rm and} 
\;\;\;\;\tau = \left [\matrix{1&1\cr -1& -1\cr}\right ] \eqno (9).
$$
In the FV formulation the equation of motion (3) takes on the form
$$
i \dot \phi = - {1\over {2m}} \nabla^2 \tau \phi + m \eta \phi + 
{g \over {2m}} \tau \phi \chi, \eqno (10)
$$
or, upon using (4), the form
$$\eqalignno {
i \dot \phi = &- {1\over {2m}} \nabla^2 \tau \phi + m \eta \phi 
+ {g \over {2m}} \tau \phi \chi_0 \cr
& + {g\over{2m}} \tau \phi \int d{\xp}\,D(x-\xp) \rho (\xp), & (11) \cr }
$$
where $\ds \rho = - g \varphi^* \varphi = - {g\over{2m}} 
\phid \eta \tau \phi$. Equation (11) is derivable from the 
Lagrangian density 
$$\eqalignno {
{\cal L}_{FV}(x) = &\;i \phi^{\dag}(x) \eta \dot \phi(x)  - 
{1\over {2m}} \nabla \phib(x) \cdot \nabla \phi (x) - m \phi^{\dag} 
(x)\phi (x) \cr & - {g \over {2m}} \phib (x) \phi (x) \chi_0 (x) + 
\oneha \int d\xp \, \rho(x) D(x-\xp) \rho(\xp), & (12) \cr}
$$
where $\phib = \phi^{\dag} \eta \tau$.
Note that ${\cal L}$ of eq. (7) is not identical to ${\cal L}_{FV}$.  
Indeed ${\cal L} = {\cal L}_{FV} + {\di \over {\di t}} (\varphi^*
{\dot \varphi})$. However, they lead to identical equations of motion 
((6) and (11)), and so are equivalent in this sense. Henceforth, 
we base our results on ${\cal L}_{FV}$.

We note that the momenta corresponding to $u$ and $v$ are 
$$
p_u = {\di {\cal L} \over {\di \dot u}} = i u^*,\;\;\; {\rm and} \;\;\; 
p_v = -i v^*,$$
that is, $u^*$ and $v^*$ are, in essence, the conjugate momenta, so 
that the Hamiltonian density is given by the expression
$$
{\cal H}(x) = \phi^{\dag}(x) \eta {\hat h} (x) \phi(x) +{g \over 
{2m}}\phib (x) \phi(x) \chi_0(x) - \oneha \int d\xp \, \rho(x) 
D(x-\xp) \rho (\xp), \eqno (13)
$$
where ${\hat h}(x) = \tau (-{1\over {2m}})\nabla^2 + m \,\eta$.

We use canonical equal-time quantization,  whereupon the non-vanishing 
commutation relations are
$$
 [u(\bx,t),p_u(\by,t)] = i \delta^N(\bx-\by)\;\;\;\; {\rm and} \;\;\;\;
[v(\bx,t),p_v(\by,t)] = i \delta^N(\bx-\by), \eqno (14)
$$
or equivalently,  
$$
[\phi_a(\bx,t),\phi^{\dag}_b(\by,t)] = \eta_{ab}\delta^N(\bx-\by),
\;\; \;\;\; a,b=1,2\eqno (15)
$$
and where $\phi^T = [\phi_1=u, \phi_2=v]$, while $\eta_{ab}$ are 
elements of the $\eta$ matrix (9).
Using these commutation relations, the QFTheoretic Hamiltonian 
can be written as
$$H = \int d^N x\, [{\cal H}_0(x) + {\cal H}_{\chi} (x) + {\cal H}_I 
(x) ], \eqno (16)$$
where
$$
{\cal H}_0 (x) = \phi^{\dag}(x) \eta {\hat h} (x) \phi(x), \eqno (17)
$$
$$ 
{\cal H}_{\chi} (x) =  {g\over{2m}}  \phib (x) \phi (x) \chi_0 (x), 
\eqno (18)
$$
and
$$\eqalignno {
{\cal H}_I (x) &=  - {g^2 \over {8m^2}} \int d\xp \, \phib (x) 
\phi(x) D(x-\xp)\,\phib (\xp) \phi (\xp) \cr
& = - {g^2 \over {8m^2}} \int d\xp \,D(x-\xp)\, \phib (x)(\phib (\xp) 
\phi(\xp))\phi(x), & (19) \cr}
$$
and where we have used $\tau^2 = 0$ in the last step of (19). Note 
that no infinities are dropped upon normal ordering, since none 
arise on account of the $\tau^2 = 0$ property.

As already mentioned, ${\cal H}_I$ contains the covariant chion 
propagator, hence in conventional covariant perturbation theory it 
leads to Feynman diagrams with internal chion lines. On the other 
hand, ${\cal H}_{\chi}$ corresponds to Feynman diagrams with external 
chions.  However, we shall not pursue covariant perturbation theory 
in this work, and so shall not consider that approach further. Rather, 
we shall consider an approach that leads to some exact eigenstates of 
the Hamiltonian (16), but with ${\cal H}_\chi = 0$. Such a truncated 
Hamiltonian is appropriate for describing states for which there is 
no annihilation or decay involving the emission or absorbtion of real chions.

In the Schr\"odinger picture we can take $t=0$. Therefore, we shall 
use the notation that, say  $\phi(\bx,t=0) = \phi(\bx)$, etc., for 
QFT operators.
This allows us to express (19) as
$$
{\cal H}_I (\bx) =  - {g^2 \over {8m^2}} \int d^N\xp \, G(\bx-\bxp) \,
\phib (\bx)(\phib (\bxp) \phi(\bxp))\phi(\bx), \eqno (20)
$$
where
$$
G(\bx-\bxp) = \int_{-\infty}^{\infty} D(x-\xp)\, dt^{\prime}= 
{1\over {(2\pi)^N}} \int d^Np\, e^{i\bp \cdot (\bx-\bxp)} {1\over 
{\bp^2 + \mu^2}}\,. \eqno(21)
$$
Explicitly, for $N=3$ this becomes
$$
G(\bx-\bxp) = {1\over {4\pi}} {e^{-\mu |\bx-\bxp|} \over {|\bx-\bxp|} }, 
\eqno (22)
$$
for $N=2$ it is
$$
G(\bx-\bxp) = {1\over {2\pi}} K_0(\mu |\bx-\bxp|),\eqno(23)
$$
where $K_0(z)$ is the modified Bessel function,
whereas for $N=1$ it has the form
$$
G(x-\xp) = {1\over {2\mu}} e^{-\mu|x-\xp|} \;. \eqno (24)
$$

At this stage we proceed in an unorthodox fashion, and
define an empty vacuum state, $|\tilde 0\rangle$, such that 
$$
\phi_a |\tilde 0\rangle = 0.  \eqno (25)
$$
This is different from the conventional Dirac vacuum $|0\rangle$ 
(the ``filled negative energy sea" vacuum), which is annihilated 
by only the positive frequency part of $\varphi$ and by the negative 
frequency part of $\varphi^*$ (see, for example, ref. [14], p. 38).  
With the definition (25), the state defined as
$$
|\psi_1\rangle = \int d^Nx\,\phid (\bx) \eta f(\bx) |\tilde 0
\rangle, \eqno (26)
$$
where f(\bx) is a two-component vector,
is an eigenstate of the truncated QFT Hamiltonian (16) (${\cal H}_\chi 
= 0$) with eigenvalue $E_1$ provided that the $f(\bx)$ is a solution 
of the equation
$$
{\hat h}(\bx) f(\bx) = E_1 f(\bx). \eqno (27)
$$
This is just the free-particle Klein-Gordon equation  for stationary 
states ($|\psi_1\rangle$ is insensitive to $H_\lambda$), in the FV 
formulation . It has, of course, all the usual negative-energy 
``pathologies'' of the KG equation. 
The presence of negative-energy solutions is a consequence of the 
use of vacuum (25). However, that is the price that has to be paid 
in order to obtain {\sl exact} eigenstates of the truncated Hamiltonian 
( eq. (16) with ${\cal H}_\chi = 0$).
We shall refer to $|\psi_1\rangle$ as a one-KG-particle state.
\vskip 1truecm
{\ni \bf  2. Two-particle eigenstates}
\vskip .5truecm
We can define two-KG-particle states, analogously to (26):
$$
|\psi_2\rangle = \int d^Nx\,d^Ny\; F_{a\,b}(\bx,\by)\;\phid_a (\bx) 
\phid_b(\by) |\tilde 0\rangle, \eqno (28)
$$
where summation on repeated indices $a$ and $b$ is implied. This state 
is an eigenstate of the truncated QFT Hamiltonian ( (16) with  ${\cal H}_
\chi = 0$) provided that the $2 \times 2$  coefficient matrix $F = 
[F_{a\,b}]$ is a solution of the two-body equation,
$$
\eta {\hat h}(\bx) \eta F(\bx,\by)  + [\eta {\hat h}(\by) \eta 
F^T(\bx,\by)]^T  
+ V(\bx-\by) \tau^T F(\bx,\by) \tau  
= E_2 F(\bx,\by), \eqno  (29)
$$
where the superscript $T$ stands for ``transpose''. The potential 
here is given by
$$
V(\bx-\by) = -{g^2 \over {4m^2}} G(\bx-\by), \eqno (30)
$$
where $G$ is specified in equations (21)-(24).  Equation (29) is a 
relativistic two-body Klein-Gordon-Feshbach-Villars-like equation, 
with an attractive Yukawa interparticle interaction.
If $V =0$, then eq. (29) has the solution $F(\bx,\by)=g_1(\bx) g_2^T(\by)$, 
where each $f_i(\bx)=\eta\,g_i(\bx)$ is a solution of the free KG equation 
(27), with eigenenergy $\varepsilon_i$, where $E_2 = \varepsilon_1 + 
\varepsilon_2$, as would be expected.

 In the rest frame, ${\bf P}_{\rm total} |\psi_2\rangle = 0$,  
 equation (29) simplifies to
$$
{\tilde h}(\br) F(\br) + [{\tilde h}(\br) F^T(\br)]^T + V(\br) 
\tau^T F(\br) \tau = E_2 F(\br), \eqno (31)
$$
where $\br = \bx - \by$, ${\tilde h} = \eta {\hat h} \eta$, and 
$\ds V(\br) = -{g^2 \over {4m^2}} G(\br)$ in this case.
It is useful to write out this equation in component form, with
$$
F(\br) = \left [ \matrix {s(\br) & t(\br)\cr u(\br) & v(\br)\cr} 
\right ], \eqno (32)
$$
namely
$$
-{1\over {2m}}\nabla^2(2s -u -t) + V (s-t-u+v) + (2m - E_2) s = 0, 
\eqno (33)
$$
$$
-{1\over {2m}}\nabla^2(s - v) + V (s-t-u+v) - E_2 t = 0, \eqno (34)
$$
$$
-{1\over {2m}}\nabla^2(s - v) + V (s-t-u+v) - E_2 u = 0, \eqno (35)
$$
and
$$
-{1\over {2m}}\nabla^2(t+u-2v) + V (s-t-u+v) - (2m + E_2) v = 0. \eqno (36)
$$
Equations (34) and (35) imply that $t(\br) = u(\br)$, so that only three 
equations survive:
$$
(2m - E_2 - \kem + V) s + (\kem - 2V) t + V v = 0, \eqno (37)
$$
$$
(-\kemm + V) s - (E_2 + 2V) t + (\kemm + V) v = 0, \eqno (38) 
$$
and
$$
V s + ( - \kem - 2 V) t - (2m + E_2 - \kem - V) v = 0. \eqno (39)
$$
These equations have positive-energy solutions of the type $E_2 = 
m+m+\cdots$, negative-energy solutions of the type $E_2=-m-m+\cdots$, 
and ``mixed'' type solutions with $E_2=m-m+\cdots$ (this is clear, 
for example, if $V=0$ and the particles are at rest). 

For the positive-energy solutions, if we write $E_2=2m+\epsilon$, then 
in the non-relativistic limit $|(\epsilon + V + {\ds {p^2\over m}} )v| 
\ll |m v|$ (and similarly for $s$ and $t$), and so equation (38) and 
(39) show that $t$ and $v$ are small and doubly-small components 
respectively, by factors $O\left ({\ds{\epsilon\over m}}\right )$. 
Thereupon, equation (37) reduces to
$$
-\kem s(\br) + V(\br) s(\br) = \epsilon s(\br), \eqno (40)
$$
which is the usual time-independent Schr\"odinger equation for 
the relative motion of two particles, each of mass $m$, interacting 
through the potential $\ds V(\br)$. Similarly, in the non-relativistic 
limit, $v$ is the large component for the negative-energy solutions 
( i.e. $E_2 = - (2m+\epsilon)$, and $s \to v$, $V \to -V$ in (40)), 
while $t$ is the large component for the mixed  energy solutions. 
 This is obvious from the form of the free-particle solutions ($V=0$), 
 which are
$$
F(\br) = s_0 \left [\matrix{  1& ({p\over{\omega+m}})^2\cr ({p\over{
\omega+m}})^2& ({{\omega-m}\over{\omega+m}})^2\cr } \right ] 
e^{i\bp\cdot\br} \;\;\;\;  
{\longrightarrow \atop {{p\over m} \ll 1}}
\;\;\;\; s_0 \left [ \matrix{  1 & ({p\over {2m}})^2   \cr 
({p\over {2m}})^2 & ({p\over {2m}})^4  \cr  } \right ] 
e^{i\bp\cdot\br}, \eqno (41)
$$
for $E_2 = 2 \omega = 2 \sqrt {p^2+m^2}$,
$$
F(\br) = t_0 \left [\matrix{{p^2\over{2m^2+p^2}}& 1 \cr 
1 & {p^2\over{2m^2+p^2}}\cr}\right ]e^{i\bp\cdot\br}, \eqno (42)
$$
for $E_2 = 0$, and
$$
F(\br) = v_0 \left [\matrix{ ({{\omega-m}\over{\omega+m}})^2& 
({p\over{\omega+m}})^2\cr ({p\over{\omega+m}})^2& 1\cr}\right ]
e^{i\bp\cdot\br}\;\;\;\;
{\longrightarrow \atop {{p\over m} \ll 1}}
 \;\;\;\; v_0 \left [\matrix{({p\over {2m}})^4& ({p\over {2m}})^2\cr 
 ({p\over {2m}})^2& 1\cr}\right ]e^{i\bp\cdot\br}, \eqno (43)
$$
for $E_2 = - 2 \omega = - 2 \sqrt {p^2+m^2}$, and where $s_0, t_0$ 
and $v_0$ are constants.

 Equations (37)-(39) can be reduced by taking suitable linear 
 combinations, whereupon it follows that ($E = E_2$)
$$
(2m +E)v = (2m-E)s + 2 E t \eqno (44)
$$
and
$$
[E(4m^2-E^2) - 8 m^2 V] s = - [E(2m+E)^2 + 8 m^2 V]t . \eqno (45)$$
It is easily verified that the free particle solutions (41) - (43), 
in particular, satisfy these relations.  One can therefore write
$$
s = [8m^2V+E(2m+E)^2]w \eqno (45)
$$ 
and
$$
t = [8m^2V+E(E^2-4m^2)]w, \eqno (46)
$$
where $w$ is a solution of ($E \ne 0$)
$$
- {1\over m} \nabla^2 w + {m\over {E/2}} V w = {1\over m} (({E/2})^2 - m^2) 
w. \eqno (47)
$$
Once (47) is solved for $w$, the components $s,t=u,v$ of the 
matrix $F$ follow from (45) and (46).

Equation (47) is form-identical to the Schr\"odinger equation, 
and so can be solved in the same manner as the latter for both 
bound and continuum states. In general this has to be done numerically. 
In some cases, such as for $\mu = 0$ (massless chion exchange) in 3+1, 
and for $\mu \ne 0$ in 1+1, analytic solutions of Schr\"odinger's 
equation are known.  We shall not discuss solution of equation (47) 
for the entire range of the parameters $\mu$ and $g$, for various $N$. 
Rather, we shall consider bound states for the two analytically 
solvable cases in some detail, leaving most of the rest for another time.  
We shall also examine bound states for one 3+1 case, with $\mu/m = 0.15$, 
numerically, since this case was studied in some detail recently by 
\NT  [8b].

\vskip 1truecm
{\ni \bf 3. Two-body bound states in 3+1 for massless chion exchange.}
\vskip .5truecm
We consider, first, the solution of equation (47) 
for $N=3$ and massless chion exchange (i.e. $\mu =0$). In this case 
one can use the known hydrogenic solutions of the Schr\"odinger 
equation to obtain the solutions of eq. (47).  Thus, for the bound 
states we obtain the result 
$$
({E/2})^2 - m^2 = - {{m^4 \alpha^2}\over {n^2 E^2}}, \eqno (48)
$$
where $\ds \alpha = {g^2\over{16 \pi m^2}}$, and $n=1,2,3,\cdots$ 
is the principal quantum number.
This yields the positive energy two-particle bound-state  spectrum
$$
E = m \sqrt{2 \left (1 + \sqrt {1-\aln^2}\right )} = m \left (2 - 
{1\over 4} {\aln}^2 - {5\over{64}} \aln^4 + \cdots \right )\;\;, \eqno (49)
$$
which is seen to have the correct Rydberg non-relativistic limit. 
Note that the relativistic spectrum retains the ``accidental'' 
Coulomb degeneracy with respect to ${\ell}$, unlike the Klein-Gordon 
spectrum for a static electromagnetic potential. This may seem surprising at first 
glance, but upon reflection it is not so. The potential $\ds V = -e A^0 
= -{\alpha \over r}$ enters both linearly and quadratically for the KG 
equation in an external electromagnetic field $A^\mu = (A^0,{\bf 0})$, 
whereas in the present  scalar Yukawa theory the potential enters only linearly . 

The two-particle bound-state wave functions corresponding to the 
eigenenergies (49) can be  lifted similarly from the Schr\"odinger 
hydrogenic results. For example, 
the ground state wave function (unnormalized) corresponding to (49) 
with $n=1$ is $\ds w = e^{-\beta r}$, where $\ds  \beta = {{m^2 \alpha}
\over E}$. The critical value of $\alpha$ beyond which $E$ (eq. (49)) 
ceases to be real is, evidently, $\alpha = 1$ ($\alpha = n$, in general).

	The question arises how these positive-energy bound-state 
	solutions compare to corresponding results obtained in 
	other formulations of this model.  The original Wick-Cutkosky 
	solutions [1,2] of the massless chion case in the ladder 
	Bethe-Salpeter approximation, as well as the corresponding 
	light-cone formulation [4,5], give the small-$\alpha$ 
	expansion
$$
E = m \left (2 - {\alpha^2 \over {4 n^2}} - {{\alpha^3 \ln \alpha} 
\over {\pi n^2}} + O(\alpha^3)\right ). \eqno (50)
$$
This is different from the present results (49), for which the 
lowest-order correction to the Rydberg energy is $O(\alpha^4)$.  
The unusual $\alpha^3 \ln \alpha$ and 
$\alpha^3$ terms (which have been termed a ``disease'' of the 
ladder Bethe-Salpeter Wick-Cutkosky solution [15]) are, apparently, 
an artifact of the {\sl ladder} approximation.  Variational-perturbative calculations 
of the scalar Yukawa model [9] that use a conventional (Dirac 
``filled negative-energy sea'') vacuum, also yield no $O(\alpha^3 
\ln \alpha, \alpha^3)$ terms. However the $O(\alpha^4)$ terms of 
[9] are different from those of  equation (49). For example, for 
the ground state [9] give
 $$
 E = m \,\left (2- {1\over 4} \alpha^2 + {{19} \over {64}} \alpha^4 
 + \cdots \right ), \eqno (51)
$$
and there is no $\ell$-degeneracy for states with $n > 1$. The 
disagreement between (51) and (49) at $O(\alpha^4)$ may be, in part, 
because virtual annihilation effects, which contribute at $O(\alpha^4)$, 
have not been included in the calculations of ref. [9] (i.e. eq. (51)).  
We might mention that a comparison of corresponding results for the 
Coulomb QED model (i.e. QED in the Coulomb gauge with the transverse 
${\bf {\alpha \cdot  A}}$ interaction turned off), namely calculations 
analogous to the present that use an ``empty'' vacuum [16], and 
conventional-vacuum variational-perturbative results [17] are in 
agreement at $O(\alpha^4)$. Thus, the reason for the lack of agreement 
between (49) and (51) beyond $O(\alpha^2)$ in the perturbative domain 
($\alpha \ll 1$) is, at present, not clear to the author and is a 
matter that needs further investigation. The disagreement with the 
ladder Bethe-Salpeter results (50) beyond $O(\alpha^2)$ is  
less surprising, given the oft-mentioned shortcomings of the {\sl ladder} 
Bethe Salpeter approximation (see, for example, the discussion in refs. 
[8b], [15], [21], [24]).

It might be tempting to speculate that the reason for the disagreement in 
the perturbative domain lies buried in the use of the unconventional 
vacuum (25). However, as has already been mentioned, the use of such 
a vacuum in the Coulomb QED case, leads to no disagreement up to 
$O(\alpha^4)$.  The present approach (with an empty vacuum) seems to 
correspond to 
the summation of a particular subset of diagrams in the conventional 
perturbative treatment (namely ladder and crossed-ladder diagrams; see also
sect. 5).  Such has been shown to be the case for the 
model theory where a (second) quantized Dirac field interacts with a 
classical (c-number) electromagnetic field [30].  
However, since a detailed analysis along the lines of [30] has not been 
carried out for the scalar Yukawa model, this remains a point of 
speculation at this stage.

In the non-perturbative regime, the present results (which are not 
perturbative) decrease monotonically with increasing $\alpha$ 
to $E= \sqrt{2}\, m$ at $\ds {\alpha \over n} = 1$. This behaviour 
is characteristic of  calculations of $E(\alpha)$ for this model in 
that $E(\alpha)$ decreases monotonically from $E(\alpha=0)=2 m$ in 
all cases. However, the various  approaches yield results that differ 
markedly in detail. For example, for the ground state ($n=1$), the 
present results decrease much more rapidly with increasing $\alpha$ 
than any of the ladder Bethe-Salpeter calculations [2,4,5,6,8] or 
the Haag-expansion results of Raychaudhuri [18]. In particular, none 
of the latter give a restriction $0 \le \alpha \le 1$ as does the present 
analytic formula (49), and so no critical value of $E(\alpha=1) = 
\sqrt {2}\,m$. Indeed the ladder Bethe-Salpeter and the variational
-perturbative approximations all give values of $E/m$, for $n=1$, 
which are near or above $1.9$ at $\alpha=1$, far above the present 
value of $1.414\cdots$, though the variational-perturbative results 
fall below the ladder Bethe-Salpeter ones for strong coupling. 
(A comparative plot of the ladder Bethe-Salpeter and variational-
perturbative values is given in fig. 1 of ref. [9]).  Raychaudhuri's 
Haag-expansion results for $\mu =0$ are also very different in detail 
from the present analytic results (see table II of ref. [18]), however 
they do exhibit a critical value of $\alpha$, but at a much larger 
value of $\alpha_c^{\rm Ray} \simeq 2.31$ (versus $\alpha_c =1$ in 
our case) with the corresponding value of $E_c^{\rm Ray} = 1.30$ 
(vs. $1.414\cdots$ here).

There are no negative-energy bound state solutions in the present 
system since, as has been mentioned earlier, the potential 
effectively reverses sign for the negative-energy case (this 
also happens in the case of one-particle relativistic equations, 
such as the Klein-Gordon-Coulomb and Dirac-Coulomb equations). 
However, eq. (48) has ``mixed-energy'' type  solutions with
$$
E = m \sqrt{2 \left (1 - \sqrt{1- \aln^2}\right )} = m \left 
({\alpha \over n} + {1\over 8} \aln^3 + {7\over{128}} \aln^5 
+ \cdots \right )\;\;. \eqno (52)
$$
These unphysical solutions do not have a Rydberg non-relativistic 
limit, and arise because of the retention of negative-energy solutions 
in the present formalism.
For these mixed-energy solutions $E$ increases monotonically with 
increasing $\alpha$ from a value of $E=0$ at $\alpha=0$ to the value 
$E/m = \sqrt{2}$ at $\alpha=n$. It is of interest to note that the 
positive-energy and mixed-energy solutions join smoothly at $\alpha=n$. 
Thus, for $0 \le \alpha < \alpha_c$, $E(\alpha)$ forms a continuous 
double-valued function, with the upper branch being the positive-energy 
solution and the lower branch being the mixed-energy solution. 
This is clear if one notes that eq. (48) can be recast into the 
equation of the semicircle $\ds {\left({E^2 \over {2m^2}}-1\right)^2 
+ \left({\alpha \over n}\right)^2 = 1}$,  $\alpha \ge 0$. Thus, eqs. 
(49) and (52) correspond, respectively, to the upper and lower branches 
of this semicircle.
Solutions that include such ``mixed-energy'' behaviour arise in some 
other formulations of the relativistic two-body system. In particular 
Raychaudhuri's Haag-expansion results exhibit this ``double-valued'' 
behaviour (see fig. 5 of ref. [18]), although he does not identify the 
lower (in energy) branch as a ``mixed-energy'' phenomenon.

\vskip 1truecm
{\ni \bf 4. Two-body bound states in $1+1$.}
\vskip .5truecm
Equation (47) in 1+1 corresponds to a one-dimensional Schr\"odinger 
equation with an exponential potential. This happens to be one of the 
not numerous cases for which the bound state eigenvalues can be expressed 
in terms of common analytic functions [19]. Thus, for the present case, 
the bound state eigenvalues are given by
$$
J^{\prime}_\nu(\gamma) = 0\;\;\;\;\; {\rm even\; parity,\; and} 
\;\;\;\;\; J_{\nu}(\gamma) = 0 \;\;\;\;\; {\rm odd\; parity}, \eqno (53)
$$
where $J_\nu(\gamma)$ is the usual Bessel function, while $\nu = 2 
\sqrt{m^2- (E/2)^2}/\mu$ and $\gamma = g /\sqrt{\mu^3 E}$. 
We evaluated some solutions of equation (53) using the Maple programme. 
Their general behaviour is similar to the expressions (49) and (52), 
in that, for given $m$ and $\mu$ there are positive-energy solutions, 
for which $E/m$ starts from 2 when $g=0$ and decreases monotonically 
with increasing $g$ for $0 \le g \le g_c$, as well as mixed-energy 
solutions for which $E/m$ start from 0 at $g=0$ and increases 
monotonically with increasing $g$ for $0 \le g \le g_c$. The two 
curves become coincident at $E(g_c)$, as was the case for the 3+1 
solutions (49) and (52).  A sample of the positive-energy 
ground-state solutions of (53), for $m/\mu = 6.944$, is given 
in table 1. We also list, for the ground state,  corresponding 
non-relativistic results  as well as perturbative and variational 
results obtained previously for the real scalar Yukawa model in 1+1 
(i.e. real phions exchanging real chions) [19]. The results for  $E$ 
are listed in the table for various values of the parameter $\lambda 
= g/(4 \sqrt{\pi} m)$ .  This parameter is not dimensionless in 1+1, 
unlike in 3+1 (where $\alpha = \lambda^2$ is dimensionless).  The 
present relativistic results are seen to fall increasingly below the 
variational and perturbative results as $\lambda$ increases. This is 
reminiscent of what was observed in 3+1.
Also, the variational results (like the perturbative and non-relativistic 
ones) give no indication of a critical value of $\lambda_c = g_c/(4 
\sqrt{\pi} m)$ beyond which no real solution for $E$ is obtained from (53).

 The behaviour of the first few even-parity excited-state solutions of 
 eq. (53) is qualitatively similar to that of the ground state. Table 2 
 is a list of the critical values of $E(\lambda_c)$ for the ground and 
 first three even-parity excited states, for $m/\mu = 6.944$.

The mixed-energy solutions of eq. (53) for the ground state of the 
$m/\mu = 6.944$ case are listed in table 3. We have not done an 
investigation of the spectrum of energies for the entire range of 
values of $m/\mu$ in 1+1, as these can be readily obtained from eq. 
(53) as needed.  


\vskip 1truecm
{\ni \bf 5. Two-body bound states in 3+1 for massive chion exchange.}
\vskip .5truecm
\NT recently reported an interesting study of the $\varphi^2 \chi$ 
model using a Feynman - Schwinger formulation [8b] that contains all 
ladder and crossed ladder diagrams.  They give a detailed comparison of 
their results with ladder Bethe-Salpeter and various ``quasipotential'' 
equations 
(i.e. modifications of the Bethe-Salpeter equation), among them the 
Blankenbecler-Sugar equation [20], the Gross equation [21] and the 
``equal-time equation'' [22,23].  Since these various equations are written 
out and discussed in ref. [8b] we shall not repeat this here. \NT find 
that the ladder Bethe-Salpeter results increasingly underestimate the 
two-body binding energy, as $\alpha$ increases, compared to their 
Feynman-Schwinger ladder-plus-crossed-ladder results, and, indeed, 
compared to the various quasipotential results (see fig. 1 of ref. [8b], 
in which $E/m$ are plotted for $0 \le \alpha < 0.93$). The various 
quasipotential results are distributed between the ladder Bethe-Salpeter 
and the \NT  Feynman-Schwinger results.  

It is of interest, therefore, to compare the predictions of the 
present approach with those given in the work of \NT [8b]. Thus, 
we calculated the positive-energy $E/m$ values for various $0 \le 
\alpha \le \alpha_c$ for this $\mu/m = 0.15$ case. We used the Maple 
differential equation solver and the ``shooting method'' (see, for 
example, ref. [25] sect. 2.5). We tested the accuracy of this 
numerical procedure on the analytic cases discussed above in sections 
3 and 4, and found that the analytic results could be reproduced with 
essentially arbitrary accuracy (we tried up to 10 figures for some 
cases).  Since the results of the various approaches compared by \NT, 
as well as ours, converge at small $\alpha$ (to the non-relativistic 
values), and since they all exhibit the same monotonically decreasing 
shape, we present here a list of $E/m$ only at $\alpha = 0.9$ where 
the differences are quite marked. These results, in decreasing order 
of $E/m$, are: Ladder Bethe-Salpeter  1.963, Blankenbecler-Sugar  1.931,  
Gross 1.910, present results of eq. (47) 1.882, Gross (with retardation) 
1.880, equal-time 1.860 and \NT 1.770. These values, except for the 
present calculation, were read off Fig. 1 of \NT [8b] and so the quoted 
accuracy of the last figure is doubtful.  A more detailed list of our 
results for this $\mu/m = 0.15$ case is given in Table 4, alongside the 
\NT and Gross-equation results (taken from Fig.1 of [8b]).

A pictorial representation of most of these positive-energy solutions is 
given in Fig. 1. We plot our present results (solid line), along with the 
Ladder Bethe-Salpeter (dashed curve), the Blankenbecler-Sugar (diamonds), 
Gross with retardation (crosses), equal-time (stars) and \NT (circles) 
values. All these results, save ours, are taken from Fig. 1 of \NT [8b]. 

As can be seen, our results are most similar to those of Gross (with 
retardation). As such, they are substantially below the Ladder 
Bethe-Salpeter results, and close (or closer) to those with ladder and 
cross ladder diagrams. This suggests that the present formalism is 
equivalent, in some sense, to a summation of all ladder and crossed 
ladder diagrams in covariant perturbation theory. The fact that the 
present formalism is non-perturbative and contains the chion propagator 
as the effective potential, is supportive of this conjecture. However, 
it would be necessary to do an analysis like that of Guiasu and Koniuk 
[30] to demonstrate this explicitly and convincingly.

It is interesting to note that the Gross results exhibit the 
``double-valued'' structure of $E(\alpha)$, with an upper 
``positive-energy'' and a lower ``mixed-energy'' branch that join 
continuously at $\alpha_c^{\rm G}$ (see inset in figure 1 of \NT [8b]). 
This is precisely the behaviour that arises in the present formalism 
(e.g. eqs. [49] and [52] for $\mu = 0$).  Indeed, the Gross critical 
value of $E(\alpha_c^{\rm G} =  1.28) \simeq 1.4 m$ is quite close to 
the value that we obtain, namely $E(\alpha_c = 1.2087) = 1.48386 m$. 
Our mixed-energy results for $\mu/m=0.15$ are also similar to those 
obtained from the Gross eq., though our results lie somewhat higher. 
For example, we obtain $E(\alpha=0.25)=0.219, E(\alpha=0.5)=0.446,  
E(\alpha=1.0)=0.985$, whereas the Gross-eq. results are $E^{\rm G}
(\alpha=0.25)=0.07, E^{\rm G}(\alpha=0.5)=0.36, E^{\rm G}(\alpha=1.0)
=0.83$.

Table 5 contains a list of our $E(\alpha)$ for $\mu/m=0.15$ in 3+1 for 
the two lowest excited states (labelled $n=2$) with  $\ell = 0$ (radial 
wave-functions $w(r)$ with one node) and  $\ell =1$ (nodeless $w(r)$). 
The general shape of $E(\alpha)$ for these excited states is similar to 
that of the ground state, except that the critical values of $\alpha$ 
(which are similar, but not coincident for these $n=2$ states) are over 
twice as large as for the ground state.  This is reminiscent of the 
$\alpha_c = n$ behaviour of the massless chion case ($\mu =0$).  There 
are mixed-energy branches of $E(\alpha)$ for the excited states, just 
as for the ground state (and the analytic $\mu =0$ case), however we do 
not give a list of them here. Their behaviour is very similar to the case 
already discussed, in that $E(\alpha)$ increases monotonically from 
$E(\alpha=0)=0$ to $E(\alpha_c)$.

\vfill \eject 
{\ni \bf 6. Three-body eigenstates}
\vskip .5truecm
It is straightforward to write down three-body states analogous to 
the two-body state (28), namely
$$
|\psi_3\rangle = \int d^Nx_1\, d^Nx_2\,d^Nx_3\;F_{abc}(\bx_1,\bx_2,\bx_3)
\,\phi_a^{\dag}(\bx_1) \phi_b^{\dag}(\bx_2) \phi_c^{\dag}(\bx_3)|\tilde 
0\rangle. \eqno (54)
$$
These are eigenstates of the Hamiltonian (16) (with ${\cal H}_\chi = 0$) 
corresponding to the eigenvalue $E_3$ provided that the $2^3=8$ coefficient 
functions $F_{abc}(\bx_1,\bx_2,\bx_3)$ are solutions of 
$$
\eqalignno{
{\tilde h}&_{ak}(\bx_1) F_{kbc}(\bx_1,\bx_2,\bx_3) +
{\tilde h}_{bk}(\bx_2) F_{akc}(\bx_1,\bx_2,\bx_3) +
{\tilde h}_{ck}(\bx_3) F_{abk}(\bx_1,\bx_2,\bx_3)\cr
&+ V(\bx_1-\bx_2) {\tilde \tau}_{ak_1} {\tilde \tau}_{bk_2} F_{k_1k_2c}
(\bx_1,\bx_2,\bx_3)
 + V(\bx_2-\bx_3) {\tilde \tau}_{bk_1} {\tilde \tau}_{ck_2} F_{ak_1k_2}
 (\bx_1,\bx_2,\bx_3) & (55)\cr
&+ V(\bx_3-\bx_1) {\tilde \tau}_{ck_1} {\tilde \tau}_{ak_2} F_{k_1bk_2}
(\bx_1,\bx_2,\bx_3)
= E_3 F_{abc}(\bx_1,\bx_2,\bx_3)\;.\cr}
$$  
Equations (55) are relativistic three-body Klein-Gordion-Feshbach-Villars-like
 equations. They are similar in structure to three-fermion equations 
derived previously [11, 26, 27], except that $\tilde h$ is then the Dirac 
operator and $a,b,c,...$ are Dirac spinor indices, so that there are 
$4^3 = 64$ equations (not all of them independent, however).
Solution of the equations (55) is much more challenging than for the 
two-body case, as it is for any relativistic three-body problem [28, 29, 31]. 
We shall not discuss the solution of these equations here.  Generalizations 
for $N$-body eigenstates can be written down in an analogous fashion.

\vskip 1truecm
\ni {\bf Concluding remarks}
\vskip .5truecm
We have shown that the scalar Yukawa model can be recast in a form such 
that exact few-body eigenstates of the QFTheoretic Hamiltonian, in the 
canonic equal-time formalism, can be determined for the case where there 
are no free (physical) quanta of the mediating field (i.e. only virtual 
quanta).  This is achieved by the partial elimination of the mediating 
field by means of Green functions, as well as by the use of the 
Feshbach-Villars formulation of scalar FT and the use of an ``empty'' 
vacuum state. This last requirement leads to the retention of 
negative-energy solutions, akin to the one-particle Klein-Gordon and 
Dirac equations.  

We considered the solution of the resulting two-particle equations in 
some detail, for massless and massive quantum exchange, in 1+1 and 
3+1 dimensions.  Analytic solutions for the two-particle bound state 
eigenenergies were obtained for massive exchange in 1+1 (eq. 53), and 
for massless exchange in 3+1 (eq. 49).
	For the massive exchange case in 3+1 our results compare 
	favourably with recent  covariant Bethe-Salpeter based models, 
	which include ladder and crossed-ladder diagrams ( particularly 
	those of the Gross equation). However, our results give stronger
 two-body binding energies than results which contain only ladder diagrams 
(such as the ladder Bethe-Salpeter results), but weaker binding than results 
that go beyond ladder-plus-crossed-ladder effects (such as the Feynman-
Schwinger calculations of \NT).

The present approach can be used for relativistic three-body systems, 
and we derived such equations for the present scalar Yukawa model.  
It can also be used for other QFT models, such as spinor Yukawa model
 and QED.

\vskip 1truecm
\ni {\bf Acknowledgement}
\vskip .5truecm
	The financial support of the Natural Sciences and Engineering 
	Research Council of Canada for this work is gratefully acknowledged.

\vskip 1truecm

{\ni \bf References}
\medskip
\par

\item{1.} G. C. Wick, Phys. Rev. {\bf 96}, 1124 (1954).
\par\noindent
\item{2.} R. E. Cutkosky, Phys. Rev. {\bf 96}, 1135 (1994).
\par\noindent
\item{3.} E. E. Salpeter and H. A. Bethe, Phys. Rev. {\bf 84}, 1232 (1951); 
H. A. Bethe and E. E. Salpeter, Phys. Rev. {\bf 87}, 328 (1952).
\par\noindent
\item{4.} G. Feldman, T. Fulton and J. Townsend, Phys. Rev. D {\bf 7}, 
1814 (1973).
\par\noindent
\item{5.} C.-R. Ji and R. J. Funstahl, Phys. Lett. {\bf 167B}, 11 (1986).
\par\noindent
\item{6.} C. Schwartz, Phys. Rev.  {\bf 137}, B717 (1965); {\bf 141}, 
1454 (1966).
\par\noindent
\item{7.} G. B. Mainland  and J. R. Spence,  Few-Body Systems 
{\bf 19}, 109 (1995).
\par\noindent
\item{8.} T. Nieuwenhuis and J. A. Tjon, Few-Body Syst. {\bf 21}, 
167 (1996); Phys. Rev. Lett. {\bf 77}, 814 (1996).
\par\noindent
\item{9.} L. Di Leo  and J. W. Darewych,  Can. J. Phys. {\bf 70}, 412 (1992).
\par\noindent
\item{10.} J. W. Darewych, Annales de la Fondation Louis de Broglie, in press, 1996.
\par\noindent
\item{11.} J. W. Darewych,  {\sl Interparticle Interactions and Nonlocality in QFT },
 in {\sl Causality and Locality in Modern Physics}, G. Hunter, J.-P. Vigier, S. Jeffers
 (eds.), Kluwer Ac. Pub., Dordrecht, The Netherlands, 1998.
\par\noindent
\item{12.} H. Feshbach and F. Villars, Rev. Mod. Phys.  {\bf 30}, 24 (1958).
\par\noindent
\item{13.} J. W. Darewych, {\sl Exact eigenstates and `triviality' of 
$\lambda (\varphi^* \varphi)^2$ theory in the Feshbach-Villars formulation}, 
York Univ. preprint, Apr. 1997  and hep-ph/9704330, and Phys. Rev, D 
{\bf 56}, 8103 (1997).
\par\noindent
\item{14.} J. D. Bjorken and S. D. Drell, {\sl Relativistic Quantum Fields} 
(McGraw Hill, New York, 1965).
\par\noindent
\item{15.} C. Itzykson and J.-B. Zuber, {\sl Quantum Field Theory} 
(McGraw Hill, New York, 1980).
\par\noindent

\item{16.} J. W. Darewych and L. Di Leo, J. Phys. A: Math. Gen. 
{\bf 29}, 6817 (1996).
\par\noindent

\item{17.} J. W. Darewych and M. Horbatsch, J. Phys. B {\bf 22}, 
973 (1989); {\bf 23}, 337 (1990).
\par\noindent

\item{18.} A. Raychaudhuri, Phys. Rev. D {\bf 18}, 4658 (1978).
\par\noindent

\item{19.} J. W. Darewych, D. V. Shapoval, I, V, Simenog and 
A. G. Sitenko, {\sl Nonperturbative formulation of relativistic 
two-particle states in the scalar Yukawa model}, J. Math. Phys. 
{\bf 38}, 3908  (1997).
\par\noindent

\item{20.} R. Blankenbecler and R. Sugar, Phys. Rev. {\bf 142}, 1051 (1966).
\par\noindent

\item{21.} F. Gross,  Phys. Rev. C {\bf 26}, 2203 (1982).
\par\noindent

\item{22.} S. J. Wallace and Y. B. Mandelzweig, Nucl. Phys. {\bf A 503}, 
673 (1989); N. K. Devine and S. J. Wallace, Phys. Rev, C {\bf 51}, 
3223 (1995).
\par\noindent

\item{23.} E. Hummel and J. A. Tjon, Phys. Rev. C {\bf 49}, 21 (1994); 
P. C. Tiemeijer and J. A. Tjon,  Phys. Rev. C {\bf 49}, 494 (1994).
\par\noindent

\item{24.} R. M. Woloshyn and A. D. Jackson, Nucl. Phys, {\bf B 64}, 
269 (1973).
\par\noindent

\item{25.} M. Horbatsch, {\sl Quantum Mechanics using Maple}, 
Springer Verlag, Berlin, 1996.
\par\noindent

\item{26.} J. W. Darewych, Phys. Lett. {\bf A 147}, 403 (1990).
\par\noindent

\item{27.} W. C. Berseth and J. W. Darewych, Phys. Lett. {\bf A 158}, 
361 (1991); {\bf A 178}, 347 (1993) and Erratum {\bf A 185}, 503 (1994).
\par\noindent

\item{28.} L. Di Leo and J. W. Darewych, Can. J. Phys. {\bf 71}, 365 (1993).
\par\noindent

\item{29.} F. Gross,  Phys. Rev. C {\bf 26}, 2218 (1982).
\par\noindent

\item{30.} I. Guiasu and R. Koniuk, Can. J. Phys. {\bf 71}, 360 (1993).
\par\noindent

\item{31} J. Bijtebier, Nucl. Phys {\bf A 508}, 305 (1990).
\vfill \eject


 \noindent 
Table 1. Values of $E_2$ (ground state) for $m=6.944$ (units: $\mu=1$) 
in 1+1.
\vskip .5cm
\vbox{\offinterlineskip
\hrule
\halign{&\vrule#&\strut\quad\hfil#\quad\cr
height10pt&\omit&&\omit&&\omit&&\omit&&\omit&\cr
&$\lambda$\hfil&&$E_2$\hfil&&$E_2$\hfil&&$E_2$ [19]\hfil&&$E_2$ [19]\hfil
&\cr
&$\displaystyle ={g\over{4\sqrt{\pi}m}}$&& eq. (53)\hfil&& non-rel.\hfil
&& variational \hfil&&pert.\hfil&\cr
height10pt&\omit&&\omit&&\omit&&\omit&&\omit&\cr
\noalign{\hrule}
height10pt&\omit&&\omit&&\omit&&\omit&&\omit&\cr
&0.0\hfill&&13.888\hfill&&13.888\hfill&&\hfill13.888\hfill&&13.888\hfill&\cr
height5pt&\omit&&\omit&&\omit&&\omit&&\omit&\cr
&0.05\hfill&&$13.8866\;8289$\hfill&&$13.8866\;8318$\hfill&&\hfill13.888
\hfill&&13.887\hfill&\cr
height5pt&\omit&&\omit&&\omit&&\omit&&\omit&\cr
&0.1\hfill&&$13.8746\;2814$\hfill&&$13.8746\;5453$\hfill&&\hfill13.879
\hfill&&13.875\hfill&\cr
height5pt&\omit&&\omit&&\omit&&\omit&&\omit&\cr
&0.2\hfill&&$13.7885\;3427$\hfill&&$13.7898\;3995$\hfill&&\hfill13.805
\hfill&&13.791\hfill&\cr
height5pt&\omit&&\omit&&\omit&&\omit&&\omit&\cr
&0.4\hfill&&$13.2754\;9678$\hfill&&$13.3205\;9729$\hfill&&\hfill13.384
\hfill&&13.334\hfill&\cr
height5pt&\omit&&\omit&&\omit&&\omit&&\omit&\cr
&0.6\hfill&&$12.0275\;4474$\hfill&&$12.4161\;0158$\hfill&&\hfill12.578
\hfill&&12.472\hfill&\cr
height5pt&\omit&&\omit&&\omit&&\omit&&\omit&\cr
&0.7\hfill&&$10.7258\;6117$\hfill&&$11.7945\;8644$\hfill&&\hfill12.578
\hfill&&12.472\hfill&\cr
height5pt&\omit&&\omit&&\omit&&\omit&&\omit&\cr
&0.75\hfill&&$9.2844\;6172$\hfill&&$11.4408\;8286$\hfill&&\hfill12.578
\hfill&&12.472\hfill&\cr
height5pt&\omit&&\omit&&\omit&&\omit&&\omit&\cr
&$0.7589\;10218^*$\hfill&&$8.2861\;8251$\hfill&&$$\hfill&&\hfill&&\hfill&\cr
height5pt&\omit&&\omit&&\omit&&\omit&&\omit&\cr
&0.8\hfill&&\hfill&&$11.0583\;6849$\hfill&&\hfill11.404\hfill&&\hfill11.201
\hfill&\cr
height5pt&\omit&&\omit&&\omit&&\omit&&\omit&\cr
&1.0\hfill&&\hfill&&$9.2380\;4702$\hfill&&\hfill9.887\hfill&&9.528\hfill&\cr
height5pt&\omit&&\omit&&\omit&&\omit&&\omit&\cr
&1.2\hfill&&\hfill&&$6.9492\;3608$\hfill&&\hfill8.057\hfill&&7.454\hfill&\cr
height5pt&\omit&&\omit&&\omit&&\omit&&\omit&\cr
&1.4\hfill&&\hfill&&$4.1878\;1169$\hfill&&\hfill5.940\hfill&&4.985\hfill&\cr
height5pt&\omit&&\omit&&\omit&&\omit&&\omit&\cr
&1.6\hfill&&\hfill&&$0.9507\;0057$\hfill&&\hfill3.561\hfill&&2.122\hfill&\cr
height5pt&\omit&&\omit&&\omit&&\omit&&\omit&\cr
&1.8\hfill&&\hfill&&$-2.7644\;9139$\hfill&&\hfill0.942\hfill&&\hfill&\cr
height5pt&\omit&&\omit&&\omit&&\omit&&\omit&\cr}

\hrule}
$^* \lambda_c$
\vskip 1truecm

\noindent 
Table 2. Values of $E_2(\lambda_c)$ in 1+1 for $m=6.944$ (units: $\mu=1$) 
for the ground state (labelled n=1) and first three even-parity excited 
states (labelled n=3,5,7).
\vskip .5cm
\vbox{\offinterlineskip
\hrule
\halign{&\vrule#&\strut\quad\hfil#\quad\cr
height10pt&\omit&&\omit&&\omit&&\omit&\cr
&$n$\hfil&&$\lambda_c$\hfil&&$E_2(\lambda_c)$\hfil&&$E_2(\lambda_c)/m$
\hfil&\cr
height10pt&\omit&&\omit&&\omit&&\omit&\cr
\noalign{\hrule}
height10pt&\omit&&\omit&&\omit&&\omit&\cr
&1\hfill&&$0.7589\;1021\;83$\hfill&&8.2862\hfill&&\hfill1.1933\hfill&\cr
height5pt&\omit&&\omit&&\omit&&\omit&\cr
&3\hfill&&$1.0429\;3738\;8$\hfill&&8.9174\hfill&&\hfill1.2842\hfill&\cr
height5pt&\omit&&\omit&&\omit&&\omit&\cr
&5\hfill&&$1.2762\;9938\;2$\hfill&&9.3428\hfill&&\hfill1.3454\hfill&\cr
height5pt&\omit&&\omit&&\omit&&\omit&\cr
&7\hfill&&$1.4983\;1853\;8$\hfill&&9.6932\hfill&&\hfill1.4348\hfill&\cr
height5pt&\omit&&\omit&&\omit&&\omit&\cr}

\hrule} 
\vfill \eject
\noindent 
Table 3. Ground-state mixed-energy solutions $E_2(\lambda)$ in 1+1 for 
$m=6.944$ (units: $\mu=1$). 
\vskip .5cm
\vbox{\offinterlineskip
\hrule
\halign{&\vrule#&\strut\quad\hfil#\quad\cr
height10pt&\omit&&\omit&\cr
&$\lambda$\hfil&&$E_2(\lambda)$\hfil&\cr
height10pt&\omit&&\omit&\cr
\noalign{\hrule}
height10pt&\omit&&\omit&&\omit&\cr
&$0.1018\;5940\;83$\hfill&&0.1&\cr
height5pt&\omit&&\omit&\cr
&$0.1440\;4069\;17$\hfill&&$0.2$&\cr
height5pt&\omit&&\omit&\cr
&$0.2276\;3468\;26$\hfill&&$0.5$&\cr
height5pt&\omit&&\omit&\cr
&$0.3213\;4930\;72$\hfill&&1.0&\cr
height5pt&\omit&&\omit&\cr
&$0.4511\;8881\;61$\hfill&&2.0&\cr
height5pt&\omit&&\omit&\cr
&$0.6192\;0329\;79$\hfill&&4.0&\cr
height5pt&\omit&&\omit&\cr
&$0.7179\;6575\;40$\hfill&&6.0&\cr
height5pt&\omit&&\omit&\cr
&$0.7455\;6999\;26$\hfill&&7.0&\cr
height5pt&\omit&&\omit&\cr
&$0.7582\;2397\;98$\hfill&&8.0&\cr
height5pt&\omit&&\omit&\cr}
\hrule}
\vskip 1truecm
\noindent 
Table 4. Values of $E_2/m$ (ground state) for $\mu/m=0.15$ in 3+1. 
\vskip .5cm
\vbox{\offinterlineskip
\hrule
\halign{&\vrule#&\strut\quad\hfil#\quad\cr
height10pt&\omit&&\omit&&\omit&&\omit&&\omit&&\omit&\cr
&$\alpha$\hfil&&Niewenhuis\hfil&&Equal-time\hfil&&Gross eq. \hfil&&Present 
\hfil&&Gross eq.\hfil&\cr
&\hfill&&and Tjon [8b]\hfil&&ref. [8b]\hfil&& (with retard.) 
\hfil&&results\hfil&&ref. [8b]\hfil&\cr
height10pt&\omit&&\omit&&\omit&&\omit&&\omit&&\omit&\cr
\noalign{\hrule}
height10pt&\omit&&\omit&&\omit&&\omit&&\omit&&\omit&\cr
&0.3\hfill&&\hfill&&\hfill&&\hfill&&$1.999\,536$\hfill&&\hfill&\cr
height5pt&\omit&&\omit&&\omit&&\omit&&\omit&&\omit&\cr
&0.4\hfill&&\hfill$1.99$\hfill&&\hfill&&\hfill&&$1.995\,34$\hfill&&\hfill&\cr
height5pt&\omit&&\omit&&\omit&&\omit&&\omit&&\omit&\cr
&0.5\hfill&&\hfill$1.98$\hfill&&\hfill&&\hfill&&$1.986\,30$\hfill&&\hfill&\cr
height5pt&\omit&&\omit&&\omit&&\omit&&\omit&&\omit&\cr
&0.6\hfill&&\hfill$1.96$\hfill&&\hfill$1.966$\hfill&&\hfill1.969\hfill&
&$1.971\,76$\hfill&&\hfill1.974\hfill&\cr
height5pt&\omit&&\omit&&\omit&&\omit&&\omit&&\omit&\cr
&0.7\hfill&&\hfill$1.91$\hfill&&\hfill$1.941$\hfill&&\hfill1.948\hfill&
&$1.950\,81$\hfill&&\hfill1.959\hfill&\cr
height5pt&\omit&&\omit&&\omit&&\omit&&\omit&&\omit&\cr
&0.8\hfill&&\hfill$1.85$\hfill&&\hfill$1.907$\hfill&&\hfill1.919\hfill&
&$1.921\,99$\hfill&&\hfill1.938\hfill&\cr
height5pt&\omit&&\omit&&\omit&&\omit&&\omit&&\omit&\cr
&0.9\hfill&&\hfill$1.77$\hfill&&\hfill$1.861$\hfill&&\hfill1.880\hfill&
&$1.882\,82$\hfill&&\hfill1.910\hfill&\cr
height5pt&\omit&&\omit&&\omit&&\omit&&\omit&&\omit&\cr
&1.0\hfill&&\hfill&&\hfill&&\hfill&&$1.828\,47$\hfill&&\hfill&\cr
height5pt&\omit&&\omit&&\omit&&\omit&&\omit&&\omit&\cr
&1.1\hfill&&\hfill&&\hfill&&\hfill&&$1.746\,64$\hfill&&\hfill&\cr
height5pt&\omit&&\omit&&\omit&&\omit&&\omit&&\omit&\cr
&1.2\hfill&&\hfill&&\hfill&&\hfill&&$1.562\,48$\hfill&&\hfill&\cr
height5pt&\omit&&\omit&&\omit&&\omit&&\omit&&\omit&\cr
&1.205\hfill&&\hfill&&\hfill&&\hfill&&$1.534\,20$\hfill&&\hfill&\cr
height5pt&\omit&&\omit&&\omit&&\omit&&\omit&&\omit&\cr
&1.208\hfill&&\hfill&&\hfill&&\hfill&&$1.503\,66$\hfill&&\hfill&\cr
height5pt&\omit&&\omit&&\omit&&\omit&&\omit&&\omit&\cr
&1.2085\hfill&&\hfill&&\hfill&&\hfill&&$1.492\,83$\hfill&&\hfill&\cr
height5pt&\omit&&\omit&&\omit&&\omit&&\omit&&\omit&\cr
&$1.2087^*$\hfill&&\hfill&&\hfill&&\hfill&&$1.483\,86$\hfill&&\hfill&\cr
height5pt&\omit&&\omit&&\omit&&\omit&&\omit&&\omit&\cr
}\hrule}
$^* \alpha_c$
\vskip 1truecm
%
%
\noindent
Table 5. Values of $E_2/m$ in 3+1 for $\mu/m=0.15$ 
for the first two excited states.
\vskip .5cm
\vbox{\offinterlineskip
\hrule
\halign{&\vrule#&\strut\quad\hfil#\quad\cr
height10pt&\omit&&\omit&&\omit&&\omit&\cr
&$\alpha$\hfil&&$E_2/m\;\;(n=2,\ell=0)$\hfil&&$\alpha$\hfil&&$E_2/m\;\;
(n=2,\ell=1)$\hfil&\cr
height10pt&\omit&&\omit&&\omit&&\omit&\cr
&\hfil&&(one node)\hfil&&\hfil&&(no nodes)\hfil&\cr
height10pt&\omit&&\omit&&\omit&&\omit&\cr
\noalign{\hrule}
height10pt&\omit&&\omit&&\omit&&\omit&\cr
&\hfill1.0\hfill&&\hfill$1.999\,954$\hfill&&\hfill1.4\hfill&&\hfill
$1.999\,031$\hfill&\cr
height5pt&\omit&&\omit&&\omit&&\omit&\cr
&\hfill1.25\hfill&&\hfill$1.996\,28$\hfill&&\hfill1.5\hfill&&\hfill
$1.995\,22$\hfill&\cr
height5pt&\omit&&\omit&&\omit&&\omit&\cr
&\hfill1.5\hfill&&\hfill$1.985\,75$\hfill&&\hfill2.0\hfill&&\hfill
$1.951\,8$\hfill&\cr
height5pt&\omit&&\omit&&\omit&&\omit&\cr
&\hfill2.0\hfill&&\hfill$1.937\,1$\hfill&&\hfill2.5\hfill&&\hfill
$1.846\,1$\hfill&\cr
height5pt&\omit&&\omit&&\omit&&\omit&\cr
&\hfill2.5\hfill&&\hfill$1.822\,7$\hfill&&\hfill2.8\hfill&&\hfill
$1.686\,2$\hfill&\cr
height5pt&\omit&&\omit&&\omit&&\omit&\cr
&\hfill2.7\hfill&&\hfill$1.725\,7$\hfill&&\hfill2.82\hfill&&\hfill
$1.663\,9$\hfill&\cr
height5pt&\omit&&\omit&&\omit&&\omit&\cr
&\hfill2.8\hfill&&\hfill$1.625\,3$\hfill&&\hfill2.84\hfill&&\hfill
$1.635\,2$\hfill&\cr
height5pt&\omit&&\omit&&\omit&&\omit&\cr
&\hfill2.82\hfill&&\hfill$1.579\,5$\hfill&&\hfill2.85\hfill&&\hfill
$1.616\,1$\hfill&\cr
height5pt&\omit&&\omit&&\omit&&\omit&\cr
&\hfill2.825\hfill&&\hfill$1.556\,8$\hfill&&\hfill$2.86$\hfill&&\hfill
$1.589\,2$\hfill&\cr
height5pt&\omit&&\omit&&\omit&&\omit&\cr
&\hfill$2.827^*$\hfill&&\hfill$1.534\,8$\hfill&&\hfill$2.866^*$\hfill
&&\hfill$1.558\,1$\hfill&\cr
height5pt&\omit&&\omit&&\omit&&\omit&\cr}

\hrule} 
$^* \alpha_c$
\vfill \eject
\bye